\documentclass[10pt,journal,compsoc]{IEEEtran}

\ifCLASSOPTIONcompsoc
  \usepackage[nocompress]{cite}
\else
  \usepackage{cite}
\fi
\ifCLASSINFOpdf
  \usepackage[pdftex]{graphicx}
\else
  \usepackage[dvips]{graphicx}
\fi
\usepackage{amssymb}
\usepackage[ruled,vlined,linesnumbered]{algorithm2e}
\usepackage{multirow}
\usepackage{comment}
\usepackage{booktabs}

\usepackage{amsmath}
\usepackage{url}
\begin{document}
%
\title{Attention on Global-Local Representation Spaces in Recommender Systems}
%
%
%
%

\author{Munlika Rattaphun,
        Wen-Chieh Fang,
        and Chih-Yi Chiu
\IEEEcompsocitemizethanks{\IEEEcompsocthanksitem M. Rattaphun, W. C. Fang, and C. Y. Chiu are with the Department of Computer Science and Information Engineering, National Chiayi University, Taiwan, R.O.C. \protect\\
Corresponding author: Wen-Chieh Fang. \protect\\
E-mail: munlika.r@gmail.com; peteragent.ai@gmail.com; cychiu@mail.ncyu.edu.tw}
}

%
%

\markboth{Journal of \LaTeX\ Class Files}%
{Shell \MakeLowercase{\textit{et al.}}: Bare Advanced Demo of IEEEtran.cls for IEEE Computer Society Journals}
%



\IEEEtitleabstractindextext{%
\begin{abstract}
In this study, we present a novel clustering-based collaborative filtering (CF) method for recommender systems.
Clustering-based CF methods can effectively deal with data sparsity and scalability problems.
However, most of them are applied to a single representation space, which might not characterize complex user-item interactions well.
We argue that the user-item interactions should be observed from multiple views and characterized in an adaptive way.
To address this issue, we leveraged the global and local properties to construct multiple representation spaces by learning various training datasets and loss functions.
An attention network was built to generate a blended representation according to the relative importance of the representation spaces for each user-item pair, providing a flexible way to characterize diverse user-item interactions.
Substantial experiments were evaluated on four popular benchmark datasets.
The results show that the proposed method is superior to several CF methods where only one representation space is considered.
\end{abstract}

}

\maketitle

\IEEEdisplaynontitleabstractindextext

%
\IEEEpeerreviewmaketitle

\ifCLASSOPTIONcompsoc
\IEEEraisesectionheading{\section{Introduction}\label{sec:introduction}}
\else
\section{Introduction}
\label{sec:introduction}
\fi

%
%
%
%
\IEEEPARstart{R}{ecommender} systems are getting popular in many web applications.
Because customers have various opinions and preferences, satisfying their requirements becomes an important issue for content and product providers.
There is a great need for personalized recommender systems that suggest appropriate content and products to a user in a wide variety of online services, such as entertainment \cite{chen2019matching}, travel \cite{zheng2020memory}, and social network \cite{zhao2020caper}.

The general approaches used in recommender systems include collaborative filtering (CF) and content-based filtering.
The CF approach builds a model from a user's past behavior (items that are purchased, viewed, selected, or rated), as well as similar behaviors by other users.
The model is then used to recommend unseen items that will likely interest the user.
The content-based filtering approach uses a series of discrete, previously-tagged characteristics of an item to recommend other items with similar properties.
A modern recommender system often combines these two approaches into a hybrid system.
In this study, we focus on the CF approach.

Matrix factorization is one of the most popular CF techniques \cite{LiangEtal:CoFactor} \cite{HeEtal:17Neural}\cite{HsiehEtal:17Collaborative}.
The relation between users and items are represented as an interaction matrix, and the matrix is then factorized into a user latent matrix and an item latent matrix.
The similarity between the user and item is usually measured by the inner product or Euclidean distance between their latent vectors.
Missing values in the interaction matrix can thus be inferred by the predicted user-item similarity and be used to make recommendations to a user for non-interacted items.
Another widely used CF technique is the neighborhood model \cite{koren2008factorization} \cite{luo2013boosting}.
It selects the nearest neighbors for each user/item based on their similarities.
The missing values of the user/item are predicted by combining its neighbors' weighted similarities.

As the system scales with rapid growth in the numbers of users and items, the interaction proportion among the whole set of users and items becomes very small.
Then these CF techniques must deal with the data sparsity and scalability problems.
One remedy is clustering-based recommendation 
\cite{sarwar2002recommender}\cite{rana2014evolutionary}\cite{liao2016clustering}\cite{chen2019n2vscdnnr}\cite{KangMcAuley:19Candidate}.
This approach groups similar users or items into subsets by means of clustering.
From the view of clusters, interactions between the user and item subsets are no more sparse.
In addition, the clustering technique can be integrated with the index structure to overcome the scalability issue. 
In other words, instead of exhaustive search, cluster pruning can be used to keep a few candidates for evaluation with respect to a given query.
The complexity is reduced to the sub-linear time of the number of items, making recommendation more efficient.
Therefore, the clustering-based approach has been extensively developed recently to address sparsity and scalability problems.

Another key point is that most existing methods are applied to a single representation space.
However, user-item interaction behaviors are often diverse and complex; it is beneficial to observe them from multiple views and to characterize them in an adaptive way.
For example, in the clustering-based approach, we can extract not only local features from clusters but also global features from the whole dataset.
These features may reflect significant properties of various user-item interactions.
Moreover, they can be combined to derive a more discriminating representation.

Following the above discussion, we propose a novel method for clustering-based CF recommender systems that involves leveraging global and local views.
Multiple representation spaces are learned from the training data of global and local user-item interactions with different loss functions.
To fuse the information from different representation spaces, an attention network is adopted to formulate an attentive representation for each user-item pair by blending the relative weights among the representation spaces.
Compared with existing attention-based methods that use an attention model at the item level (considering which item is more important), the proposed attention model is used at the representation level (considering which representation is more important).

We highlight the contributions of the proposed method as follows:
\begin{itemize}
  \item Multiple representation spaces are constructed based on different global/local properties and loss functions.
  Rather than using only one representation as in most existing methods, multiple representations can provide an abundance of features to describe the relation between users and items.
  \item An attention network is presented to capture the relative importance of multiple representation spaces.
  It can dynamically fuse multiple representations for each user-item pair, forming a flexible measure to characterize diverse user-item interactions.
  \item Substantial experiments compared with various configurations and state-of-the-arts are evaluated under four benchmark datasets.
  The results can support the effectiveness and superiority of the proposed method.
\end{itemize}

The remainder of this paper is organized as follows.
Section~\ref{sec:related} presents related work on recommender systems.
Section~\ref{sec:method} elaborates the proposed method, including the global/local representation networks and attention network.
Section~\ref{sec:experiments} demonstrates and discusses experimental results.
Finally, conclusions are summarized in Section~\ref{sec:conclusion}.

\section{Related Work}
\label{sec:related}
Among the mass of literature studies in recommender systems, we focus on the up-to-date CF-based methods with brief introduction.
We also review the clustering-based approach and attention mechanism highly related to our work.

\subsection{CF-based recommender systems}
The core idea of the CF approach is to model the preference of users toward items according to their historical interactions.
Several techniques are proposed to address this issue.
In addition to the above-mentioned matrix factorization and neighborhood model, here we introduce other popular techniques, namely, ranking formulation and deep learning.

Rendle et al.~\cite{RendleEtal:09BPR} presented a generic optimization criterion by Bayesian personalized ranking, called BPR-Opt, for personalized ranking in the item recommendation task. 
BPR-Opt is the maximum posterior estimator derived from a Bayesian analysis of the problem. 
The main idea is to rank observed items higher than unobserved items. 
They also provided a generic learning algorithm for optimizing models with respect to BPR-Opt.
Following the similar ranking formulation, Hsien et al.~\cite{HsiehEtal:17Collaborative} proposed collaborative metric learning (CML), which encoded user preferences and user-user/item-item similarity in a joint metric space.
CML achieves significant speedup for top-\textit{N} recommendation tasks using off-the-shelf approximate nearest-neighbor search with negligible accuracy reduction.

More recent approaches tend to use the deep learning technique.
He et al.~\cite{HeEtal:17Neural} presented a general framework called neural network-based collaborative filtering (NeuMF) for recommender systems.
They used nonlinear neural networks as the user–item interaction function.
NeuMF expresses and generalizes matrix factorization and leverages a multi-layer perceptron to learn the user–item interaction function. 
The authors transformed the identity of a user/item to a binarized sparse vector with one-hot encoding, which only explored the user/item feature in a limited manner.
Due to the complexity of the scoring function, the NeuMF retrieval process is generally hard to accelerate.
Rendle et al.~\cite{RendleEtal:20Neural} revisited the NeuMF experiments, and concluded that a dot product might be a better default choice for combining embeddings than learned similarities using MLP or NeuMF.
Xue et al.~\cite{XueEtal:19Deep} proposed a neural network framework to model higher-order item relations for item-based collaborative filtering by using multiple nonlinear layers above pairwise interaction modeling.
Under this framework, the authors also used an attention model to differentiate the importance of pairwise interactions.
Deng et al.~\cite{DengEtal:19DeepCF} proposed a general framework called deep collaborative filtering (DeepCF) to combine the representation learning and matching function learning.
They also proposed a model called a collaborative filtering network (CFNet) based on a plain-vanilla MLP model under the DeepCF framework.
CFNet has great flexibility in learning complex matching functions, with good efficiency in learning low-rank relations between users and items.

To handle the high-dimensional and sparse matrix efficiently, Luo et al. presented a series studies for latent factor-based CF methods \cite{luo2021latent} \cite{luo2021algorithms} \cite{luo2021fast} \cite{wu2021deep}.
For example, they investigated the stochastic gradient descent algorithm and develop several extensions to improve the accuracy \cite{luo2021latent}.
In addition, some non-negative factorization algorithms are proposed to overcome the slow convergence problem under non-negativity constraints \cite{luo2021algorithms}\cite{luo2021fast}.
Wu et al. designed a deep structure to construct the hierarchical latent factor models that can achieve high computation and storage efficiency \cite{wu2021deep}.

\subsection{Clustering-based recommender systems}
The clustering-based approach is well known for its scalability to large and sparse systems.
The basic steps include calculating the similarity based on rating data, and then applying a clustering algorithm to group similar users and items into clusters.
Some advanced methods are introduced in the following.

Ji et al.~\cite{JiEtal:16Improving} proposed a reconstructive method for improving matrix approximation by introducing clustering and transfer learning techniques.
The proposed method compresses low-rank approximation into a cluster-level rating-pattern referred to as a codebook and then constructs an improved approximation by expanding the codebook.
Wu et al.~\cite{WuEtal:16CCCF} proposed a scalable co-clustering method called CCCF.
The idea is users may have different preferences over different subsets of items, where these subsets of items may overlap.
CCCF can explore subgroups of users who share interests over similar subsets of items through user-item co-clustering.
Chen et al.~\cite{chen2019n2vscdnnr} proposed a recommender system called N2VSCDNNR based on node2vec~\cite{grover2016node2vec} and clustering technology.
N2VSCDNNR recommends item clusters with high weights to the target user cluster.
The weights are obtained from the interaction frequency between items and users.
Then the items in the high-weight clusters are recommended to the target user. 
Kang and McAuley~\cite{KangMcAuley:19Candidate} proposed a clustering-based framework called CIGAR, which learns a preference-preserving binary embedding model and a candidate item re-ranking model.
Although using binary codes can significantly reduce the inference cost, due to their constrained capability, the performance is limited compared to models that use real-valued representations.

\subsection{Recommender systems with attention mechanisms}
The attention mechanisms are components of prediction systems that enable a system to focus sequentially on different subsets of the input~\cite{ChoEtal:15Describing}.
They have been widely adopted in recommender systems recently.

Chen et al.~\cite{ChenEtal:17Attentive} presented an attentive collaborative filtering model to address implicit feedback in multimedia recommendation.
They introduced item-level and component-level attention models to assign attentive weights for inferring the underlying user preferences encoded in implicit user feedback.
Zhu et al.~\cite{ZhuEtal:18Learning} presented a tree-based deep recommendation model that predicts user preferences using a max-heap-like tree probability formulation with an attention-based deep network.
The underlying deep model and tree structure are learned in turn.
Tay et al.~\cite{TayEtal:18Latent} proposed a metric learning approach called latent relational metric learning for recommendations.
They adopt the idea of translating users to items by translation vectors while relying on a memory attention network to learn the relation vectors.
He et al.~\cite{HeEtal:19Hierarchical} proposed a user-generated list recommendation model that leverages the hierarchical structure of items, lists, and users to capture the containment relationship between lists and items.
They applied a self-attention network to refine the item and list representations by considering the consistency of neighboring items and lists.
Hu et al. \cite{hu2021bcfnet} introduced a model named BCFNet, which is an extended version of DeepCF \cite{DengEtal:19DeepCF}.
The BCFNet consists of three sub-models, including representation learning, matching function learning, and balance module.
In particular, an attention mechanism is integrated to improve the ability of the representation learning and matching function learning.

For group recommendation, the attention mechanisms can be employed to aggregate the properties of a set of user/item group.
Cao et al.~\cite{CaoEtal:18Attentive} proposed using NeuMF with attention models to characterize user and user group interest at the same time.
They further integrated the modeling of user-item interactions into their method, making it possible to reinforce the two tasks of recommending items for users and user groups.
Chen et al. \cite{chen2019matching} considered the task of recommending a set of items (bundle) to a user.
They designed a factorized attention network to fuse the set of item representations into a bundle representation.

The attention mechanisms are also applied in sequential recommendation.
For example, Zheng et al.~\cite{zheng2020memory} proposed a memory-augmented hierarchical attention network (MAHAN) for next point-of-interest (POI) recommendation.
MAHAN incorporates two networks to tackle short-term check-in sequences and long-term memories for preference modeling.
A co-attention network is used to characterize the interaction between these long-term and short-term preferences.
Wu et al. \cite{wu2020personalized} presented a similar idea, where the long-term preference is formulated by attention weight evaluation.
Hao et al. \cite{hao2021annular} explored user's short-term preference based on the local and global features of the annular-graph of the user behavior sequence by the self-attention mechanism.
Fan et al. \cite{fan2021lighter} introduced the low-rank decomposed self-attention to generate content-aware representations, where $n$ items are aggregated into $k$ latent interests $(n > k)$ to perform a lightweight self-attention computation.

\subsection{Summary}
To sum up, the CF techniques that employ neural networks to model user-item interactions become a popular choice.
However, these techniques mainly focus on a single representation space.
We argue that it is insufficient to model the complicated user-item interactions. 
The clustering-based methods are proposed in several studies to address the sparsity and scalability problems.
Few of them take advantage of global and local representations from the cluster structure.
We point out that multiple representations are beneficial to provide multi-views for observing user-item interactions.
The attention mechanisms can provide an effective way to fuse these representations into a blended one.
We develop the representation-level attention, which is different from the item-level attention used by existing methods.

\section{Proposed Method}
\label{sec:method}
\begin{figure}[ht]
    \centering
    \includegraphics[width=.4\textwidth]{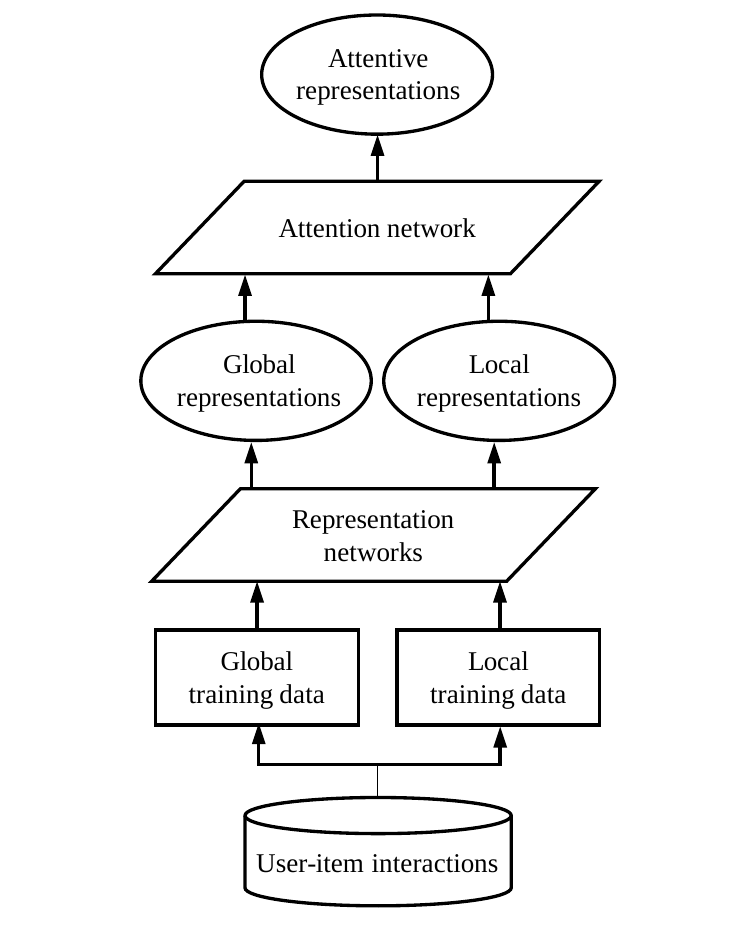}
    \caption{Overview of the training process of the proposed method.}
    \label{fig:overview}
\end{figure}

In the proposed method, we build multiple representation networks and an attention network for the recommendation purpose.
Fig. \ref{fig:overview} illustrates an overview of the training process for these networks.
Given a set of user-item interactions, we prepare two categories of training data, namely, global and local, according to whether clustering is involved or not.
Then the representation networks are trained by using different training data categories and loss functions.
Next, multiple global and local representations, which are generated from the representation networks, are employed to train the attention network.
The attention network combines these representations to obtain the attentive representations of the users and items, which are used to measure the compatibility between each user-item pair.
Details are elaborated in the following subsections.

\subsection{Global representation}
The goal of the CF approach with implicit feedback is to explore the unobserved relations between user $u$ and item $i$ and to answer the question: will the user interact with the item in the future?
A user-item interaction consists of, for example, the user assigning a score to an item or checking information about an item.
The interaction history is usually described by an interaction matrix, denoted as $A = \left [ a_{ui} \right ]$, which contains the user ID
$u$, the item ID $i$, and the interaction $a_{ui}$ that represents the preference of user $u$ for item $i$.
This study focuses on evidence that the user interacted with an item, regardless of the extent of the interaction.
Therefore, the matrix element $a_{ui}$ is defined by:
\begin{equation}
    a_{ui} =  \begin{cases} 
      1, & \text{if} \; u \; \text{interacted with} \; i, \\
      0, & \text{otherwise.}
   \end{cases}
\end{equation}
A triplet training set is sampled from $A$ and expressed as:
\begin{equation} \label{eq:global_triplet}
    T = \left \{ \left ( u, i^+, i^- \right ) \big| a_{ui^+} = 1, a_{ui^-} = 0 \right \},
\end{equation}
where $i^+$ and $i^-$ are the positive and negative items, respectively.
In practice, it is impossible to enumerate all triplet combinations and many of them are trivial for training.
We adopt a simple strategy in this study \cite{HsiehEtal:17Collaborative}\cite{pretet2020learning}.
Given the row vector of user $u$ in interaction matrix $A$, for each positive item $i^+$ (where $a_{ui^+} = 1$), we randomly choose an negative item $i^-$ (where $a_{ui^-} = 0$) to produce a triplet $(u,i^+,i^-)$.
The iteration process will repeat until the number of triplets is sufficient.

\begin{figure}[ht]
    \centering
    \includegraphics[width=.5\textwidth]{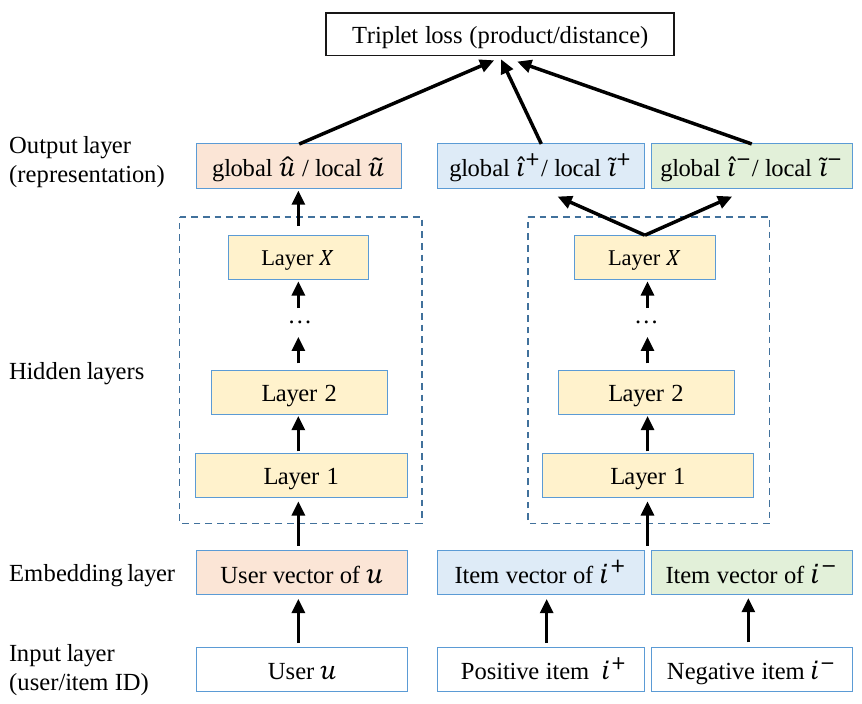}
    \caption{The representation network.}
    \label{fig:representation}
\end{figure}

To derive the global representation, local information such as cluster assignment of the training data is ignored during the training phase.
Let $G \in \mathbb{R}^d$ be a global representation space of $d$-dimensions.
To project user $u$ and item $i$ into $G$, the representation network is designed based on the Siamese network \cite{koch2015siamese}, as shown in Fig. \ref{fig:representation}.
The embedding layer transforms a user or item ID into a dense vector of a fixed size.
The embedding vector is then forwarded to hidden layers for nonlinear transformation.
Note that we modify the typical Siamese network so that the variant can accept a triplet $\left ( u, i^+, i^- \right ) \in T$ as the input in the training process.
In addition, we separate the user from the item without sharing the hidden layers.
In other words, the network parameters can be very different between the user part and the item part.
This is because users and items are different in nature, they should be processed through different paths in order to be projected to the common representation space.
Through the respective embedding and hidden layers, inputs $u$, $i^+$, and $i^-$ are transformed with L2 normalization into $G$ as representation outputs $\hat{u}$, $\hat{i}^+$, and $\hat{i}^-$, respectively.

The training of the representation network is guided by a triplet loss function.
Using different loss functions to generate variant representations can provide multiple views to characterize each user and item.
We introduce two loss functions for this purpose, the dot product loss and the Euclidean distance loss; more loss functions can be conducted if necessary.
The dot product loss originates from BPR-Opt \cite{RendleEtal:09BPR}.
Its variant adopts the conventional matrix factorization, referred to as BPR-MF, which is generally used as an underlying preference estimator.
BPR-MF can be optimized by a contrastive pairwise ranking objective function of a triplet, referred to as \textit{product loss}:
\begin{equation} \label{eq:product_loss}
    l_{prod}(\hat{u}, \hat{i}^+, \hat{i}^-) = - \ln \sigma \left ( \langle \hat{u}, \hat{i}^+ \rangle - \langle \hat{u}, \hat{i}^- \rangle \right ),
\end{equation}
where $\sigma$ is the sigmoid function and $\left \langle \cdot \right \rangle$ denotes the dot product computation.
Because enumerating all triplets in $T$ is typically intractable, BPR-MF uses stochastic gradient descent (SGD) to optimize the model.
Specifically, in each step of SGD, a batch of triplets is dynamically sampled from $T$.
In addition, L2 regularization was adopted on the user and item representations, which is crucial to alleviate overfitting.

The second global representation was generated based on the Euclidean distance metric, where the optimization function is \textit{distance loss} expressed by \cite{schroff2015facenet}:
\begin{equation} \label{eq:distance_loss}
    l_{dist}(\hat{u}, \hat{i}^+, \hat{i}^-) = max \left( 0, \| \hat{u} - \hat{i}^+ \|^2 - \| \hat{u} - \hat{i}^- \|^2 + \alpha \right),
\end{equation}
where $\alpha$ is the margin.

To make the optimization effective, hard examples of triplets that result in large loss can be selected frequently during training.
The network parameters are updated through SGD to minimize the loss function iteratively.
In the following, the global representations of user $u$ and item $i$ are denoted as $\hat{u}$ and $\hat{i}$, respectively.

\subsection{Local representation}
Roughly speaking, the global representations provide a \textit{macro} view of users and items.
However, they are insufficient to model complex user-item interactions.
The local representations, on the other hand, are derived by learning local data, which are extracted from the clusters that are close to the user.
They can model the user-item interactions from a \textit{micro} view.

For simplicity, we assume the global representation space $G$ is a Euclidean space.
\textit{K}-means clustering is used to construct a cluster set of size $M$.
Given a query (user) $u$ and its global representation $\hat{u}$, $J$ clusters are chosen from the $M$ clusters that are the closest to $\hat{u}$ and are denoted as candidate clusters $\lbrace C_j\rbrace_{j=1}^J$.
The corresponding codebook of centroids is $\lbrace \mu_j | \mu_j \in G\rbrace_{j=1}^J$.
To construct the local representation network, we employ the same network architecture defined in Fig. \ref{fig:representation} but train with the local properties of the candidate clusters.

Suppose that a positive item $\hat{i}^+$ and a negative item $\hat{i}^-$ belong to two candidate clusters, i.e., $\hat{i}^+ \in C_j$ and $\hat{i}^- \in C_k$, where $j, k \in \{ 1, \cdots, J \}$.
There are two considerations for generating the triplet training set.
The first case is $j = k$, i.e., $\hat{i}^+$ and $\hat{i}^-$ are in the same cluster.
They can be included in the \textit{local intra-triplet} set $T^\dagger \in T$ when satisfying the following criterion:
\begin{equation} \label{eq:intra-triplet}
    T^\dagger = \left \{ (u, i^+, i^-) \big| \| \hat{u}-\hat{i}^+ \| > \| \hat{u}-\hat{i}^- \| \right \}.
\end{equation}
The intuition is that, if the distance between $u$ and $i^+$ is larger than that between $u$ and $i^-$, it will incur positive triplet loss.
We can use this hard example to tune the representation network so that the user is close to its positive item and far away from its negative item.
The other case is $j \neq k$, i.e., $\hat{i}^+$ and $\hat{i}^-$ are in different clusters.
A similar definition for the \textit{local inter-triplet} set $T^\ddagger \in T$ is given:
\begin{equation} \label{eq:inter-triplet}
    T^\ddagger = \left \{ (u, i^+, i^-) \big| \| \hat{u}-\hat{\mu}_j \| > \| \hat{u}-\hat{\mu}_k \| \right \}.
\end{equation}
Unlike Eq. (\ref{eq:intra-triplet}) that considers the relation between the user and items, Eq. (\ref{eq:inter-triplet}) addresses the relation between the user and clusters.
That is, if the distance between $u$ and $C_j$ (where $i^+$ belongs to) is larger than that between $u$ and $C_k$ (where $i^-$ belongs to), we take this hard example to tune the user representation network so that the user is close to its positive cluster $C_j$ and far away from its negative cluster $C_k$.
Consequently, both $T^\dagger$ and $T^\ddagger$ serve as the local training dataset for learning local representations.

The outputs of the local representation network are expressed as the user $\tilde{u}$, the positive item $\tilde{i}^+$, and the negative item $\tilde{i}^-$.
In a similar way to Eq.~\ref{eq:product_loss} and Eq.~\ref{eq:distance_loss}, the loss functions for training local representations are expressed as $l_{prod}(\tilde{u}, \tilde{i}^+, \tilde{i}^-)$ for product loss and $l_{dist}(\tilde{u}, \tilde{i}^+, \tilde{i}^-)$ for distance loss.

\subsection{Attentive representation} \label{sec:attention}
By leveraging the global and local properties with different loss functions, multiple representation networks can be trained to generate the respective representations for each user/item.
Moreover, an attention network is proposed to find the best combination based on these representations.
That is, the proposed attention network can map a set of user-item representations to weighted sum representations of the user and item.
The weight is assigned to each representation space by computing a compatibility function of the user and item.
Details are described in the following.

\begin{figure*}[h]
    \centering
    \includegraphics[width=1.0\textwidth]{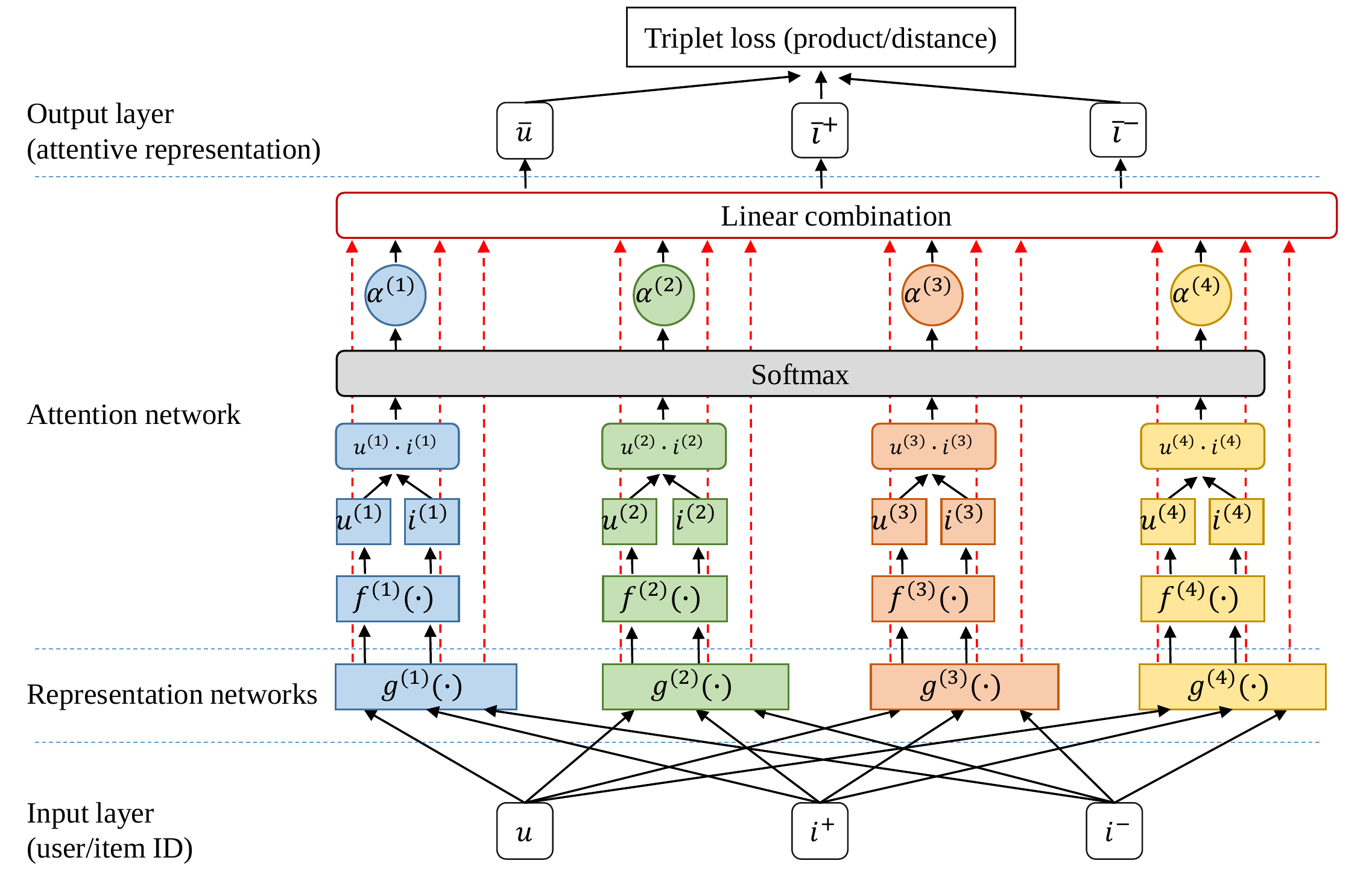}
    \caption{The proposed model consisting of $R$ representation networks followed by an attention network, where four representation spaces $(R = 4)$ are taken as an example.}
    \label{fig:attention}
\end{figure*}

Suppose that we construct $R$ representation networks to derive corresponding representation spaces; each user/item has $R$ representations.
Fig. \ref{fig:attention} gives an overview of the proposed network consisting of $R$ representation networks followed by an attention network, where four representation spaces $(R=4)$ are taken as an example.
In the training process, the input layer receives a triplet $(u, i^+, i^-)$.
Let $g^{(r)}$ be the $r$th representation network, $r \in {1, 2, ..., R}$, which transforms the user and items to their $r$th representations.
We express the $r$th representations of the triplet as $\big( g^{(r)}(u), g^{(r)}(i^+), g^{(r)}(i^-) \big)$, and forward them to the $r$th attention channel.

Each attention channel can be divided into three parts.
The first part is a transformation function, denoted as $f^{(r)}$, which maps the user and positive item to the respective vectors $u^{(r)}$ and $i^{(r)}$:
\begin{equation}
\begin{split}
    & u^{(r)} = f^{(r)} \big( g^{(r)}(u) \big), \\
    & i^{(r)} = f^{(r)} \big( g^{(r)}(i^+) \big),
\end{split}
\end{equation}
Note that the negative item is excluded in the attention network.
Consequently, the second part is to calculate the compatibility between the user and positive item.
The idea is paying more attention to the representation space where the user and the positive item are more compatible, as measured by the dot product.
The compatibility is then convert to a normalized weight, denoted as $\alpha^{(r)}$, by applying the softmax function over all representation spaces.
\begin{equation}
    \alpha^{(r)} = \frac{\exp(u^{(r)} \cdot i^{(r)})}{\sum_{r} \exp(u^{(r)} \cdot i^{(r)})}.
\end{equation}
In the last part, a weighted sum is calculated as the linear combinations of the $R$ representation spaces for each user/item:
\begin{equation} \label{eq:attention_blending}
\begin{split}
    & \bar{u} = \sum_{r=1}^{R} \alpha^{(r)} \cdot g^{(r)}(u), \\
    & \bar{i}^+ = \sum_{r=1}^{R} \alpha^{(r)} \cdot g^{(r)}(i^+), \\
    & \bar{i}^- = \sum_{r=1}^{R} \alpha^{(r)} \cdot g^{(r)}(i^-).
\end{split}
\end{equation}
The outputs $\bar{u}$, $\bar{i}^+$, and $\bar{i}^-$ are denoted as \textit{attentive representations}.
The proposed model is trained by the triplet loss function based on either product loss $l_{prod}(\bar{u}, \bar{i}^+, \bar{i}^-)$ or distance loss $l_{dist}(\bar{u}, \bar{i}^+, \bar{i}^-)$, and the training data are randomly sampled from the global and local triplet sets $T$, $T^\dagger$, and $T^\ddagger$.

\subsection{Recommendation as retrieval}
We treat the recommendation task as a retrieval process that searches for similar items from multi-view representation spaces for a given user.
In clustering-based recommendation, a coarse-to-fine search strategy can be applied without scanning the dataset exhaustively, making recommendation more efficient and scalable.
For each item $i$ in the dataset, $R$ representations are extracted by the set of representation networks $\{ {g}^{(r)}(i) | r = 1, 2, ... , R \}$.
We choose one of the global representations and perform \textit{k}-means clustering to create $M$ clusters and to build the index structure with $M$ clusters, each of which has attached an inverted list of associated items.

When a user query $u$ is submitted, the user representations $\{ {g}^{(r)}(u) \}$ are extracted and used to search for the $K$ nearest item clusters.
We denote the items stored in the nearest clusters as the candidate items.
For each pair of user $u$ and candidate item $i'$, the attention network is used to obtain the attentive representations $\bar{u}$ and $\bar{i}'$ defined in Eq.~\ref{eq:attention_blending} and to compute their Euclidean distance or dot product to estimate their similarity.
The candidates are sorted to output the top-$N$ items as the recommendation result for  user $u$.
The retrieval process is summarized in Algorithm \ref{alg:recommendation}.

By taking advantage of the coarse-to-fine search strategy, only some dataset items are selected as candidates for the attention and re-ranking computation.
Although applying the attention network for each user-item pair requires additional computations, these can be accelerated by GPUs, and thus the runtime is usually minor.
Therefore, the time complexity of the retrieval process can be considered sublinear with respect to the size of the item dataset.

\begin{algorithm}[h]
    \caption{Recommendation}
    \label{alg:recommendation}
    \KwIn{a set of $M$ item clusters, $\Phi$; the set of the corresponding $M$ inverted lists of item clusters, $\Theta$ ; a user $u$;}
    \KwOut{$N$ recommended items $\left\{ i_1, i_2, ..., i_N \right\}$;}
    Extract the user embeddings $\{ {g}^{(r)}(u) | r = 1, 2, ..., R \}$\;
    Determine the top-$K$ item clusters $\{ C_k| C_k \in \Phi \}_{k=1}^K, K<M$ for $u$ by calculating the maximum dot products or minimal Euclidean distances between $u$ and the item cluster centroid\;
    Let $I = \{i| i \in \cup_{k \in \{1,2,\ldots,K\}} L_k, L_k \in \Theta$ \} be the set of candidate items from the top-$K$ item clusters\;
    Return the top-\textit{N} items $\{ i_j|i_j \in I \}_{j=1}^N$ with the maximum dot products or minimum Euclidean distances between $\bar{u}$ and $\bar{i_j}$ from Eq.~\ref{eq:attention_blending}\;
\end{algorithm}

\section{Experimental Results}
\label{sec:experiments}
This section describes the evaluation of the proposed method and a comparison with several CF-based recommender systems.
The experiments were carried out under Windows 10 with an Intel Core i7-6700 CPU, an NVIDIA GeForce GTX 1080Ti GPU, and 64GB RAM.
The programming environment was the Keras package of Tensorflow under Python 3.7.
The source code is available at https://github.com/MunlikaRattaphun/Attention-on-Global-Local-Embedding-Spaces-in-Recommender-Systems.

\subsection{Benchmark datasets}
Four popular benchmark datasets, including MovieLens, Yelp, and Amazon, were used in the experimental evaluation.
We briefly introduce these datasets, as well as the corresponding training/test data partitions:
\begin{itemize}
    \item \textbf{MovieLens-100k.}
    This dataset consists of 100,000 records of user-movie ratings from 943 users and 1,682 movies. 
    Each user has rated at least 20 movies.
    We randomly sampled half of the user-movie views as test data, and used the other half for training.
    \item \textbf{Yelp Madison.}
    The Yelp dataset is a set of user reviews of businesses in local cities.
    Here, we selected the city of Madison.
    Users who had more than five business interactions were kept; the others were removed from the dataset.
    For each of the remaining users, we randomly sampled three user-business interactions as the test data, and used the rest for training.
    \item \textbf{Yelp Pittsburgh.}
    Yelp Pittsburgh is another dataset containing user reviews of businesses in Pittsburgh.
    The user filtering and training/test data generation were the same as for Yelp Madison.
    \item \textbf{Amazon.}
    This dataset contains product reviews from customers of Amazon.com.
    For this study, the category \textit{Movie and TV} was chosen.
    Users with fewer than six reviews were removed from the dataset.
    The training/test data were generated in the same way as for Yelp Madison and Yelp Pittsburgh.
\end{itemize}

Table \ref{tb:dataset} summarizes some statistic properties of these datasets.
Their scales and densities are very different from each other, providing a diverse way for observation.

\begin{table}
  \caption{Dataset statistics (after preprocessing)}
  \label{tb:dataset}
  \resizebox{\columnwidth}{!}{\begin{tabular}{lrrrr}
    \hline
    \textbf{Datasets} & \textbf{\#Users} & \textbf{\#Items} & \textbf{\#Interactions} & \textbf{Density} \\ 
    \hline
        MovieLens-100k    & 943              & 1,682            & 100,000       & 0.0630   \\
        Yelp Madison      & 2,773            & 3,186            & 50,961        & 0.0058   \\
        Yelp Pittsburgh   & 5,116            & 6,443            & 111,246       & 0.0034   \\
        Amazon            & 32,910           & 45,585          & 1,089,539     & 0.0007   \\
    \hline
  \end{tabular}}
\end{table}

\subsection{Evaluation metrics}
Several widely used accuracy metrics in top-\textit{N} recommendation, including precision, recall, hit rate (HR), average-reciprocal hit rank (ARHR), and normalized discounted cumulative gain (NDCG), were used for experiment evaluation.
For a user query $u$, the top-\textit{N} recommended items are denoted as $Q_u = \lbrace i_1, i_2, ..., i_N \rbrace$ and the ground truth (positive) items as $G_u = \lbrace i'_1, i'_2, ..., i'_{|G_u|} \rbrace$, where $| \cdot |$ is the set cardinality.
These metrics are defined as follows:
\begin{align}
    \text{recall} \big( u \big)  & =  \frac{ \big| Q_u \cap G_u \big|} {\big| G_u \big|}  \\
    \text{precision} \big( u \big)  & =  \frac{ \big| Q_u \cap G_u \big|} {\big| Q_u \big|}  \\
    \text{HR} \big( u \big) & =  \begin{cases} 
          1, & \text{if} \; \big| Q_u \cap G_u \big|  \geqslant 1  \\
          0, & \text{if} \; \big| Q_u \cap G_u \big| = 0 \end{cases} \\
    \text{ARHR} \big( u \big) & = \sum_{ i \in Q_u \cap G_u } \frac {1}{rank_i} \\
    \text{NDCG} \big( u \big) & = \sum_{ i \in Q_u } \frac{1}{log_2(rank_i+1)}
\end{align}
where $rank_i$ is the ranking position of item $i$.
Although a higher value reflects a better accuracy in general, it is recommended to assess the model performance from multiple metrics rather than any single metric.

\subsection{Hyperparameter and ablation study}
To study the impact of model parameters, we take the following four global and local representations, as well as two attentive representations, for consideration:
\begin{itemize}
    \item \textbf{Global representation / distance loss (GD)} was generated based on the global training set (Eq. (\ref{eq:global_triplet})) and the distance loss function (Eq. (\ref{eq:distance_loss})).
    \item \textbf{Global representation / product loss (GP)} was generated based on the global training set and product loss function (Eq. (\ref{eq:product_loss})).
    \item \textbf{Local representation / distance loss (LD)} was generated based on the local training set (Eqs. (\ref{eq:intra-triplet}) and (\ref{eq:inter-triplet})) and the distance loss function.
    \item \textbf{Local representation / product loss (LP)} was generated based on the local training set and the product loss function.
    \item \textbf{Attentive representation / distance loss (AD)} was blended from the above four representations by the proposed attention network and the distance loss function.
    \item \textbf{Attentive representation / product loss (AP)} was blended by the proposed attention network and the product loss function.
\end{itemize}
Except for using different training datasets and loss functions, these representations adopted the same representation network shown in Fig. \ref{fig:representation}, which contains one embedding layer and two hidden layers, each of which has 64 neurons.
The dimensionality of the output layer was also set to 64.
Each model was trained for a maximum of 200 epochs with an early stopping strategy and optimized with the Adam optimizer, where the batch size was fixed at 512 and the learning rate was 0.00017.
The margin of the loss function in Eq. (\ref{eq:distance_loss}) was set $\alpha = 0.5$.

\begin{figure*}[ht]
    \centering
    \includegraphics[width=1.0\textwidth]{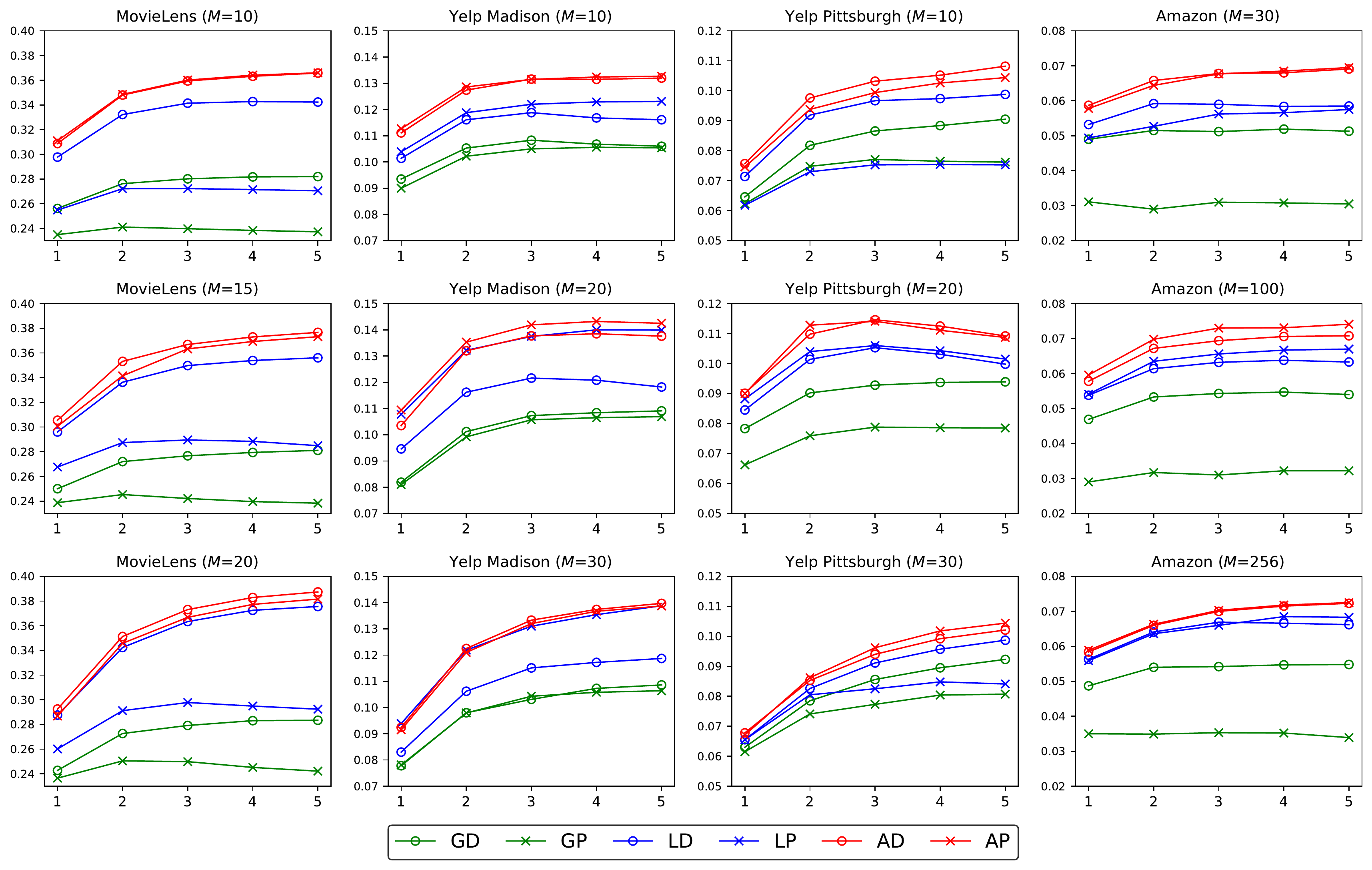}
    \caption{Top-$N$ candidate clusters (X-axis) vs. recall (Y-axis).}
    \label{fig:top_cluster_recall}
\end{figure*}

We first evaluate the impact of the number of clusters $M$.
We evaluate three different $M$ for each dataset: $M \in \lbrace 10, 15, 20\rbrace$ for MovieLens, $M \in \lbrace 10, 20, 30\rbrace$ for Yelp Madison, $M \in \lbrace 10, 20, 30\rbrace$ for Pittsburgh, and $M \in \lbrace 30, 100, 256\rbrace$ for Amazon.
The range of $M$ has been carefully chosen: too large or small $M$ will incur a bad clustering result.
The performance is plotted in Fig. \ref{fig:top_cluster_recall}, where each column shows a particular dataset with different $M$.
Each plot shows six recall curves of the proposed representations under the top-$K$ candidate clusters, $K \in \{ 1, 2, 3, 4, 5 \}$.
We ranked the items in the candidate clusters to return the top-20 items for recommendation.

As the number of candidate clusters $K$ is increased, more positive items are found from the candidate clusters, and therefore the recall curves is generally improved.
In addition, most of them show a sharp rise at $K = 2$.
However, some recall curves have a degraded performance as $K$ increases.
This means that the corresponding model does not generate good representation spaces, so that the relevance between the user and positive item is less than that between the user and negative item.
The proposed attentive representations (AD and AP) are relatively stable under these configurations, showing that they became more robust by adaptively fusing multiple representations.

In general, local representations (LD and LP) are better than global representations (GD and GP).
However, no single representation always performs better than others because of the variety of user-item interaction behaviors.
Again, the proposed attention network combining different views of representation spaces yield the best recall on all the benchmark datasets.
Based on the above result, we choose the following $M$ in subsequent experiments: $M = 20$ for MovieLens, $M = 20$ for Yelp Madison, $M = 10$ for Yelp Pittsburgh, and $M = 100$ for Amazon.

\begin{figure*}[ht]
    \centering
    \includegraphics[width=1.0\textwidth]{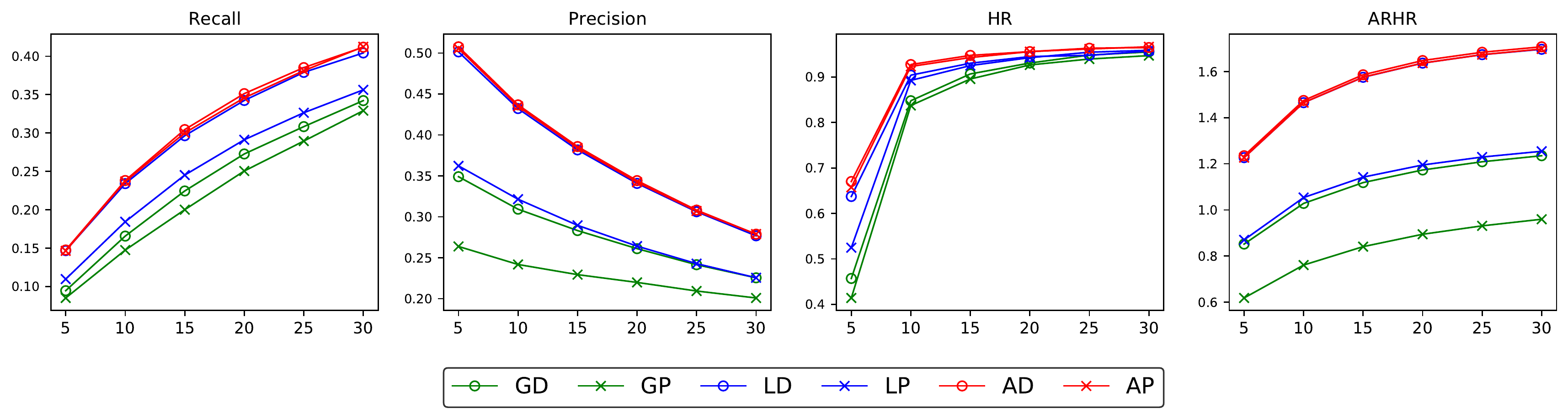}
    \caption{Top-$N$ recommended items (X-axis) vs. accuracy metrics (Y-axis) in MovieLens.}
    \label{fig:movielens}
\end{figure*}

\begin{figure*}[ht]
    \centering
    \includegraphics[width=1.0\textwidth]{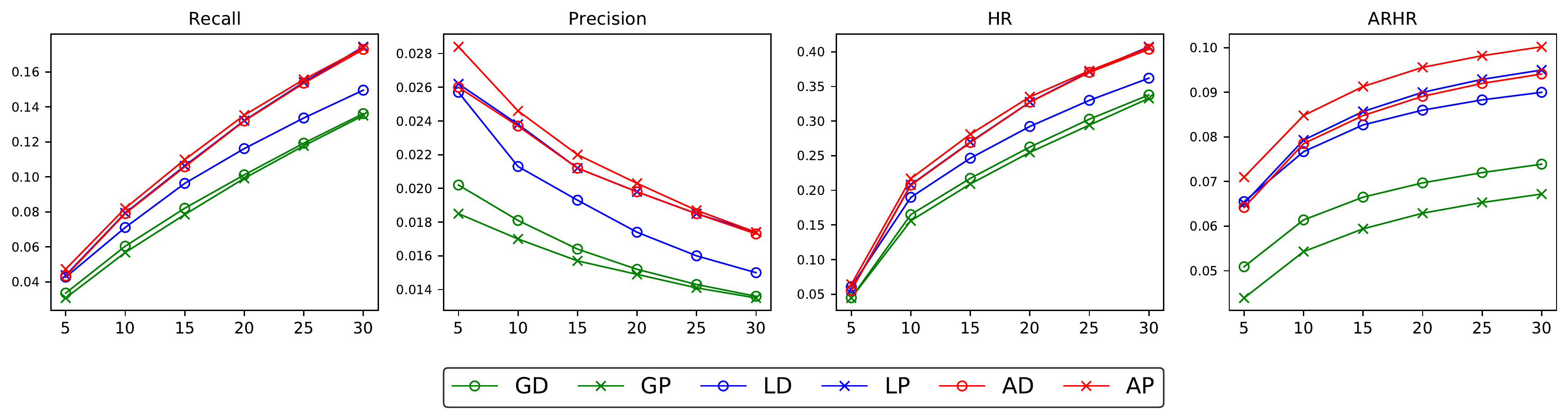}
    \caption{Top-$N$ recommended items (X-axis) vs. accuracy metrics (Y-axis) in Yelp Madison.}
    \label{fig:mcity}
\end{figure*}

\begin{figure*}[ht]
    \centering
    \includegraphics[width=1.0\textwidth]{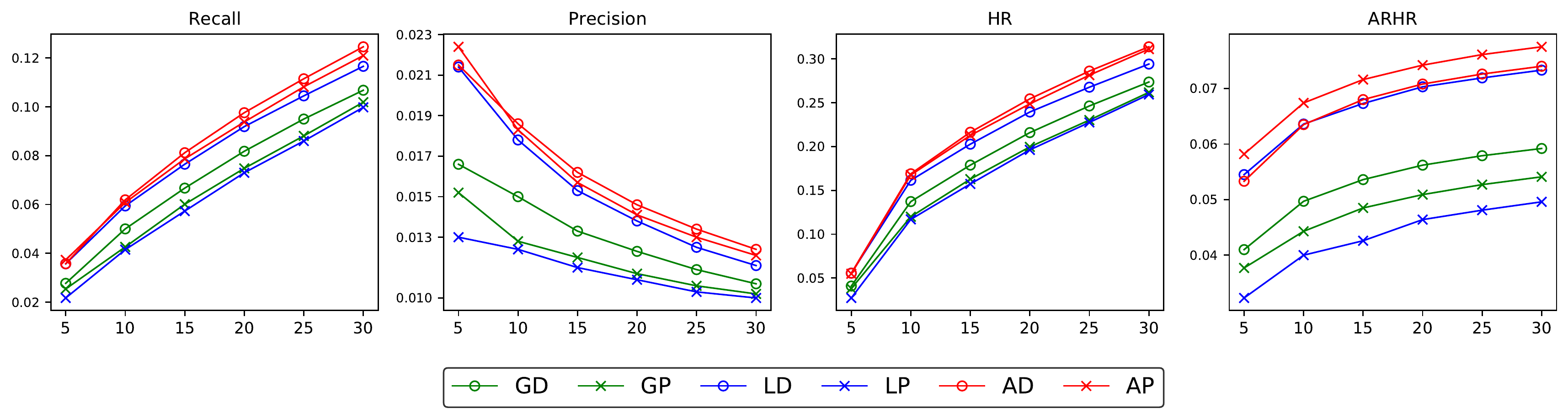}
    \caption{Top-$N$ recommended items (X-axis) vs. accuracy metrics (Y-axis) in Yelp Pittsburgh.}
    \label{fig:pcity}
\end{figure*}

\begin{figure*}[ht]
    \centering
    \includegraphics[width=1.0\textwidth]{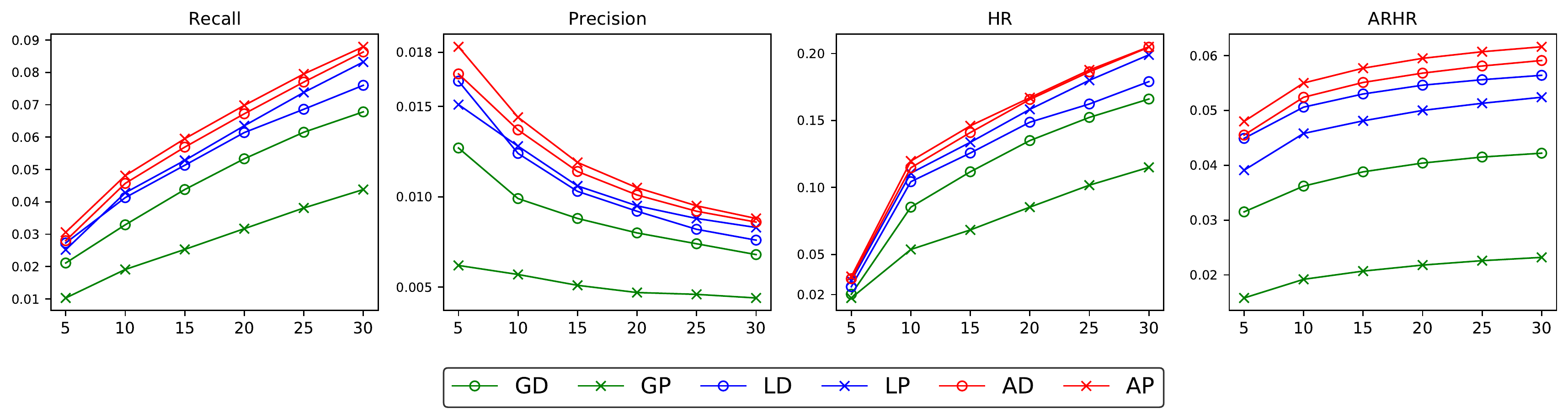}
    \caption{Top-$N$ recommended items (X-axis) vs. accuracy metrics (Y-axis) in Amazon.}
    \label{fig:amazon}
\end{figure*}

Next, we observe the recommendation result from more evaluation metrics, as shown in Figs. \ref{fig:movielens}, \ref{fig:mcity}, \ref{fig:pcity} and \ref{fig:amazon} for MovieLens, Yelp Madison, Yelp Pittsburgh, and Amazon, respectively.
Each figure contains four subplots, including recall, precision, HR, and ARHR (from left to right). The X-axis represents the number of top-$N$ recommendation items, $N \in \{5, 10, 15, 20, 25, 30\}$, and the Y-axis represents the accuracy.
The number of candidate clusters is fixed at $K = 2$.
Again we see the proposed attention networks outperform the other representation networks under these accuracy metrics.
Particularly, the proposed method AP is superior to AD under the ARHR metric, even they are comparable under the other three metrics.
Note that ARHR measures the ranking quality that assigns a high score to hits at top position ranks.
That is, AP could provide a better ranking quality.
Another interesting phenomenon is, for the HR metric, most of the curves show a sharp rise at $N = 10$.
Since HR is an indication whether at least one positive item is in the top-$N$ set, we suggest $N \geq 10$ would be a good number of recommendation items.

An ablation study is introduced for the attention network of the global and local representations.
Three representation types are considered: global, local, and all.
The global type consists of GD, GP, GAD (global attentive representations with distance loss), and GAP (global attentive representations with product loss).
The local type follows the similar definition.
We fixed the number of candidate clusters at $K = 2$ and the number of recommended items at $N = 15$.
Tables \ref{tb:ablation_1} and \ref{tb:ablation_2} list the performance in the four benchmark datasets.
The number in boldface indicates the best accuracy in each metric column.
It is clear to see that with the attention network, the global attentive representations (i.e., GAD and GAP) gain further improvement by combining the two global representations.
The local attentive representations (i.e., LAD and LAP) also gain the similar improvement.
By combining all global and local representations together, the attention network can yield the best accuracy, as demonstrated by AD and AP.

\begin{table*}
\caption{Ablation study for the attention network of the global and local representations in MovieLens and Yelp Madison}
\label{tb:ablation_1}
\centering
\begin{tabular}{l|l|cccc|cccc} 
\hline
\multicolumn{1}{l}{\multirow{2}{*}{}} & \multirow{2}{*}{} & \multicolumn{4}{c|}{MovieLens}  & \multicolumn{4}{c}{Yelp Madison}  \\ 
\cline{3-10}
\multicolumn{1}{l}{}    & & Recall    & Precision & HR    & ARHR  & Recall    & Precision & HR    & ARHR  \\ 
\hline
\multirow{4}{*}{Global type} & GD      & 0.2230    & 0.2829    & 0.9025    & 1.1060    & 0.0797    & 0.0160    & 0.2126    & 0.0671    \\
                        & GP      & 0.2001    & 0.2294    & 0.8956    & 0.8407    & 0.0790    & 0.0158    & 0.2124    & 0.0591    \\
                        & GAD    & 0.2500    & 0.2961    & 0.9275    & 1.1736    & 0.0797    & 0.0159    & 0.2124    & 0.0671    \\
                        & GAP    & 0.2391    & 0.2919    & 0.9166    & 1.1619    & 0.0818    & 0.0164    & 0.2172    & 0.0659    \\ 
\hline
\multirow{4}{*}{Local type}  & LD      & 0.2965    & 0.3817    & 0.9303    & 1.5752    & 0.0873    & 0.0174    & 0.2322    & 0.0745    \\
                        & LP      & 0.2452    & 0.2896    & 0.9251    & 1.1421    & 0.0985    & 0.0197    & 0.2563    & 0.0799    \\
                        & LAD    & 0.2958    & 0.3743    & 0.9298    & 1.5077    & 0.1008    & \textbf{0.0201} & 0.2605  & \textbf{0.0835}  \\
                        & LAP    & 0.2833    & 0.3478    & 0.9406    & 1.4750    & 0.0998    & 0.0200    & 0.2589  & 0.0804  \\ 
\hline
\multirow{2}{*}{All type}   & AD  & \textbf{0.3045} & \textbf{0.3858} & \textbf{0.9475} & \textbf{1.5861} & 0.1006    & \textbf{0.0201} & 0.2600  & 0.0829    \\
                                & AP  & 0.3002    & 0.3841    & 0.9428    & 1.5759    & \textbf{0.1009} & \textbf{0.0201} & \textbf{0.2609} & 0.0834  \\
\hline
\end{tabular}
\end{table*}

\begin{table*}
\caption{Ablation study for the attention network of the global and local representations in Yelp Pittsburgh and Amazon}
\label{tb:ablation_2}
\centering
\begin{tabular}{l|l|cccc|cccc} 
\hline
\multicolumn{2}{l|}{\multirow{2}{*}{}} & \multicolumn{4}{c|}{Yelp Pittsburgh}   & \multicolumn{4}{c}{Amazon}    \\ 
\cline{3-10}
\multicolumn{2}{l|}{}                  & Recall & Precision & HR    & ARHR  & Recall    & Precision & HR    & ARHR  \\ 
\hline
\multirow{4}{*}{Global type}       & GD     & 0.0652 & 0.0130    & 0.1786  & 0.0559  & 0.0438    & 0.0088    & 0.1117    & 0.0388    \\
                              & GP     & 0.0580 & 0.0116    & 0.1594  & 0.0464  & 0.0253    & 0.0051    & 0.0683    & 0.0207    \\
                              & GAD   & 0.0652 & 0.0130    & 0.1786  & 0.0559  & 0.0470    & 0.0094    & 0.1207    & 0.0412    \\
                              & GAP   & 0.0657 & 0.0131    & 0.1785  & 0.0549  & 0.0463    & 0.0093    & 0.1220    & 0.0412    \\ 
\hline
\multirow{4}{*}{Local type}        & LD     & 0.0710 & 0.0142    & 0.1903  & 0.0680  & 0.0513    & 0.0103    & 0.1257    & 0.0530    \\
                              & LP     & 0.0632 & 0.0127    & 0.1709  & 0.0491  & 0.0528    & 0.0106    & 0.1337    & 0.0481    \\
                              & LAD   & 0.0723 & 0.0144    & 0.1931  & 0.0682  & 0.0529    & 0.0106    & 0.1343    & 0.0482    \\
                              & LAP   & 0.0705 & 0.0141    & 0.1898  & 0.0642  & 0.0593    & \textbf{0.0119} & \textbf{0.1460} & 0.0551    \\ 
\hline
\multirow{2}{*}{All type} & AD     & 0.0722 & 0.0144    & 0.1930    & 0.0681    & 0.0569    & 0.0114    & 0.1410    & 0.0551    \\
                              & AP     & \textbf{0.0727} & \textbf{0.0145} & \textbf{0.1963} & \textbf{0.0731} & \textbf{0.0595} & \textbf{0.0119} & \textbf{0.1460} & \textbf{0.0577}  \\
\hline
\end{tabular}
\end{table*}

\subsection{Comparison with other CF methods}
We compared several CF recommendation systems, including network-based, factorization-based, and clustering-based approaches:
\begin{itemize}
    \item \textbf{MetaPath2Vec++ \cite{DongEtal:metapath2vec}:}
    This system uses a network embedding model that characterizes user-item interactions by heterogeneous network representation learning.
    \item \textbf{BiNE \cite{GaoEtal:BiNE}:}
    This system is also a network embedding model that characterizes user-item interactions by bipartite representation learning.
    \item \textbf{CoFactor \cite{LiangEtal:CoFactor}:}
    This factorization model generates the representation by jointly decomposing the user-item interaction matrix and the item-item co-occurrence matrix with shared item latent factors.
    \item \textbf{N2VSCDNNR \cite{chen2019n2vscdnnr}:}
    This clustering-based method uses the node2vec technique to generate a common representation space and builds user and item clusters for matching.
    \item \textbf{CIGAR \cite{KangMcAuley:19Candidate}:}
    This is also a clustering-based method that uses a hashing technique to retrieve candidate items and learns a ranking model to re-rank the candidate items.
\end{itemize}

Tables \ref{tb:comparison_1} and \ref{tb:comparison_2} present their performances in terms of recall, precision, HR, and ARHR.
The clustering-based approach (i.e., N2VSCDNNR, CIGAR, AP and AD) yields a better performance than the other approaches in the diverse benchmark datasets, demonstrating its superiority to address the sparsity and scalability problems to a certain extent.
Furthermore, CIGAR performs better than N2VSCDNNR in the sparser and larger datasets, i.e., Yelp and Amazon.
However, these compared methods employ a single representation, which is insufficient to model complex user-item interactions.
The proposed attention model combines multiple representations to regenerate joint representations for each user-item pair dynamically, thus gain the best performance in various benchmark datasets.

\begin{table*}
\caption{Comparison with CF methods in MovieLens and Yelp Madison}
\label{tb:comparison_1}
\centering
\begin{tabular}{l|cccc|cccc} 
\hline
\multicolumn{1}{c|}{\multirow{2}{*}{}} & \multicolumn{4}{c|}{MovieLens} & \multicolumn{4}{c}{Yelp Madison}     \\ 
\cline{2-9}
\multicolumn{1}{c|}{}                       & Recall & Precision & HR     & ARHR   & Recall & Precision & HR     & ARHR    \\ 
\hline
Metapath2vec++                              & 0.2750 & 0.2810    & 0.8957 & 1.1761 & 0.0341 & 0.0101    & 0.1214 & 0.0381  \\
BiNE                                        & 0.1236 & 0.1721    & 0.7108 & 0.7559 & 0.0542 & 0.0128    & 0.1512 & 0.0481  \\
CoFactor                                    & 0.2750 & 0.2810    & 0.8957 & 1.1761 & 0.0951 & 0.0160    & 0.2088 & 0.0657  \\
N2VSCDNNR                                   & 0.2865 & 0.2933    & 0.9350 & 1.2241 & 0.1042 & 0.0166    & 0.2166 & 0.0686  \\
CIGAR                                       & 0.1728 & 0.2687    & 0.9300 & 0.9868 & 0.1124 & 0.0225    & 0.2952 & 0.0842  \\
AD        & \textbf{0.2983} & \textbf{0.4702}    & \textbf{0.9788} & 1.9169 & \textbf{0.1355} & \textbf{0.0271}    & \textbf{0.3524} & \textbf{0.0922}  \\
AP        & \textbf{0.2983} & \textbf{0.4702}    & \textbf{0.9788} & \textbf{1.9171} & \textbf{0.1355} & \textbf{0.0271}    & \textbf{0.3524} & 0.0921  \\
\hline
\end{tabular}
\end{table*}

\begin{table*}
\caption{Comparison with CF methods in Yelp Pittsburgh and Amazon.}
\label{tb:comparison_2}
\centering
\begin{tabular}{l|cccc|cccc} 
\hline
\multicolumn{1}{c|}{\multirow{2}{*}{}} & \multicolumn{4}{c|}{Yelp Pittsburgh} & \multicolumn{4}{c}{Amazon}     \\ 
\cline{2-9}
\multicolumn{1}{c|}{}                       & Recall & Precision & HR     & ARHR   & Recall & Precision & HR     & ARHR    \\ 
\hline
Metapath2vec++                              & 0.0326 & 0.0095    & 0.1164 & 0.0426 & 0.0257 & 0.0038    & 0.0643 & 0.0223  \\
BiNE                                        & 0.0399 & 0.0115    & 0.1352 & 0.0526 & 0.0389 & 0.0059    & 0.0875 & 0.0366  \\
CoFactor                                    & 0.0735 & 0.0144    & 0.1832 & 0.0602 & 0.0575 & 0.0087    & 0.1141 & 0.0494  \\
N2VSCDNNR                                   & 0.0818 & 0.0153    & 0.2008 & 0.0644 & 0.0608 & 0.0093    & 0.1237 & 0.0528  \\
CIGAR                                       & 0.1259 & 0.0252    & 0.3155 & 0.0876 & 0.0701 & 0.0140    & 0.1789 & 0.0665  \\
AD        & \textbf{0.1345} & \textbf{0.0269}    & \textbf{0.3455} & 0.1001 & 0.0746 & 0.0149    & 0.1861 & 0.0768  \\
AP        & \textbf{0.1345} & \textbf{0.0269}    & 0.3433 & \textbf{0.1138} & \textbf{0.0793} & \textbf{0.0159}    & \textbf{0.1980} & \textbf{0.0796}  \\
\hline
\end{tabular}
\end{table*}

Finally, we performed leave-one-out evaluation to compare our methods with two other CF methods.
The compared methods are end-to-end learning-based recommender systems:
\begin{itemize}
    \item \textbf{NeuMF \cite{HeEtal:17Neural}:} This system generalizes matrix factorization by nonlinear neural networks to model user-item interactions.
    \item \textbf{BCFNet \cite{hu2021bcfnet}:} This system, which is an extension of DeepCF \cite{DengEtal:19DeepCF}, uses an attention mechanism to improve the representation learning and matching function learning.
\end{itemize}
In leave-one-out evaluation, we held-out the latest interaction as a test item for each user and utilized the remaining data for training.
For testing we followed the same way of the above two methods: we randomly selected 99 unobserved interactions as the negative examples for each user, together with the latest interaction as the positive example.
In total 100 test are ranked, and the top-10 items were evaluated by HR and NDCG.

Table \ref{tb:comparison_3} lists the leave-one-out evaluation for the four benchmark datasets.
Note the BCFNet’s result for Amazon is not available because the memory consumption is too large to be executed in our computer.
For each dataset, we generated three representation spaces of 32, 64, and 128 dimensions.
Our methods yield better HR and NDCG scores in the high-dimensional representation space.
In particular, our NDCG scores outperform NeuMF and BCFNet with a large gap, indicating our methods can rank the positive example at a superior position.
The proposed method is able to find better representations for users and items through separate representation networks.
Separating items into positive and negative ones also has an important impact on the recommendation task.
In addition, similar to BCFNet, the attention mechanism can be used to highlight important features.
However, because the local representations are derived by learning local data, which are extracted from the clusters close to the user, when the density vector is low-dimensional, it may be easier to mix some dissimilar data points in a cluster than that is high-dimensional.
The local representations that come out may be poor, and in turn affect the final recommendation results.

\begin{table}
\caption{Comparison with NeuMF and BCFNet by leave-one-out evaluation}
\label{tb:comparison_3}
\centering
\begin{tabular}{l|ll|ll|ll} 
\hline
\multirow{3}{*}{} & \multicolumn{6}{c}{MovieLens}   \\ 
\cline{2-7}
                  & \multicolumn{2}{c|}{32-d}   & \multicolumn{2}{c|}{64-d} & \multicolumn{2}{c}{128-d} \\ 
\cline{2-7}
                  & \multicolumn{1}{c}{HR} & NDCG  & \multicolumn{1}{c}{HR} & NDCG & \multicolumn{1}{c}{HR} & \multicolumn{1}{c}{NDCG}  \\ 
\hline
NeuMF             & 0.6405  & 0.3679    & 0.6617    & 0.3834    & 0.6490    & 0.3766    \\
BCFNet            & \textbf{0.6723}     & 0.3896    & 0.6702    & 0.3912    & \textbf{0.7010}   & 0.4096    \\
AD                & 0.6426  & 0.3787    & 0.7063    & 0.4675    & 0.6903    & 0.3902                    \\
AP                & 0.6501  & \textbf{0.4011}   & \textbf{0.7243}   & \textbf{0.4944}   & 0.6999    & \textbf{0.4538}   \\ 
\hline
\multirow{3}{*}{} & \multicolumn{6}{c}{Yelp Madison}    \\ 
\cline{2-7}
                  & \multicolumn{2}{c|}{32-d}   & \multicolumn{2}{c|}{64-d} & \multicolumn{2}{c}{128-d} \\ 
\cline{2-7}
                  & \multicolumn{1}{c}{HR} & NDCG   & \multicolumn{1}{c}{HR} & \multicolumn{1}{c|}{NDCG} & \multicolumn{1}{c}{HR} & \multicolumn{1}{c}{NDCG}  \\ 
\hline
NeuMF             & 0.6383  & 0.3917    & 0.6502    & 0.3986    & 0.6556    & 0.4104    \\
BCFNet            & \textbf{0.6664} & \textbf{0.4199}   & \textbf{0.6661}   & 0.4170    & 0.6682    & 0.4235    \\
AD                & 0.6304  & 0.3518    & 0.6477    & 0.4601    & 0.7010    & \textbf{0.5819}   \\
AP                & 0.6275  & 0.3357    & 0.6361    & \textbf{0.4672}   & \textbf{0.7079}   & 0.5729    \\ 
\hline
\multirow{3}{*}{} & \multicolumn{6}{c}{Yelp Pittsburgh} \\ 
\cline{2-7}
                  & \multicolumn{2}{c|}{32-d}   & \multicolumn{2}{c|}{64-d} & \multicolumn{2}{c}{128-d} \\ 
\cline{2-7}
                  & \multicolumn{1}{c}{HR} & \multicolumn{1}{c|}{NDCG} & \multicolumn{1}{c}{HR} & \multicolumn{1}{c|}{NDCG} & \multicolumn{1}{c}{HR} & \multicolumn{1}{c}{NDCG}  \\ 
\hline
NeuMF             & 0.6855  & 0.4253    & 0.6949    & 0.4324    & 0.6933    & 0.4394    \\
BCFNet            & \textbf{0.7000} & \textbf{0.4458}   & \textbf{0.7048}   & 0.4575   & 0.7062    & 0.4546    \\
AD                & 0.6571  & 0.3726    & 0.6841    & \textbf{0.5677}  & \textbf{0.7261}    & \textbf{0.6249}   \\
AP                & 0.6455  & 0.3729    & 0.6841    & \textbf{0.5677}  & 0.6754 & 0.4398    \\
\hline
\multirow{3}{*}{} & \multicolumn{6}{c}{Amazon} \\ 
\cline{2-7}
                  & \multicolumn{2}{c|}{32-d}   & \multicolumn{2}{c|}{64-d} & \multicolumn{2}{c}{128-d} \\ 
\cline{2-7}
                  & \multicolumn{1}{c}{HR} & \multicolumn{1}{c|}{NDCG} & \multicolumn{1}{c}{HR} & \multicolumn{1}{c|}{NDCG} & \multicolumn{1}{c}{HR} & \multicolumn{1}{c}{NDCG}  \\ 
\hline
NeuMF             & 0.6437  & 0.3781    & 0.6422    & 0.3774    & 0.6490    & 0.3846    \\
BCFNet            & \multicolumn{1}{c}{-} &\multicolumn{1}{c|}{-}   &\multicolumn{1}{c}{-}   &\multicolumn{1}{c|}{-}   &\multicolumn{1}{c}{-}    & \multicolumn{1}{c}{-}    \\
AD                &\textbf{0.6961}  & \textbf{0.4356}    & \textbf{0.6988}    & \textbf{0.4319}  & \textbf{0.6972}    & \textbf{0.4302}  \\
AP                & 0.6746  & 0.3770    & 0.6763    & 0.3999  &\textbf{0.6972}    & 0.4257  \\
\hline
\end{tabular}
\end{table}

\section{Conclusion}
\label{sec:conclusion}
In this study, we proposed a novel clustering-based CF method for recommender systems.
The user and item representations are learned from multiple views in terms of global/local representation spaces and dot product/Euclidean distance loss functions.
An attention network is designed to generate a joint representation of these views for each user-item pair dynamically.
Experimental results show the proposed method is effective and competitive compared to several CF methods where only one representation space is considered.


\ifCLASSOPTIONcaptionsoff
  \newpage
\fi



%




\bibliographystyle{IEEEtran}
\bibliography{bibliography}

\end{document}